\documentclass[aps,prb,twocolumn,showpacs,floatfix,superscriptaddress]{revtex4}

\usepackage{amsmath}
\usepackage{amssymb}
\usepackage{bm}
\usepackage{graphicx}

\begin{document}

\title{Localization properties of a one-dimensional tight-binding model with
non-random long-range inter-site interactions}

\author{F.\ A.\ B.\ F. de Moura}
\affiliation{Departamento de F\'{\i}sica, Universidade Federal de Alagoas,
57072-970 Macei\'{o}, AL, Brazil}

\author{A.\ V.\ Malyshev}
\thanks{On leave from Ioffe Physiko-Technical Institute, 26
Politechnicheskaya str., 194021 Saint-Petersburg, Russia}
\affiliation{Departamento de F\'{\i}sica de Materiales, Universidad Complutense,
E-28040 Madrid, Spain}

\author{M.\ L.\ Lyra}
\affiliation{Departamento de F\'{\i}sica, Universidade Federal de Alagoas,
57072-970 Macei\'{o}, AL, Brazil}

\author{V.\ A.\ Malyshev}
\thanks{On leave from S.I. Vavilov State Optical Institute, Birzhevaya Liniya
12, 199034 Saint-Petersburg, Russia.}
\affiliation{Institute for Theoretical Physics and Materials Science Center,
University of Groningen, Nijenborgh 4,  9747 AG Groningen, The Netherlands}

\author{F.\ Dom\'{\i}nguez-Adame}
\thanks{Also at Grupo Interdisciplinar de Sistemas Complejos.}
\affiliation{Departamento de F\'{\i}sica de Materiales, Universidad Complutense,
E-28040 Madrid, Spain}

\date{\today}

\begin{abstract}

We perform both analytical and numerical studies of the one-dimensional
tight-binding Hamiltonian with stochastic uncorrelated on-site energies and
non-fluctuating long-range hopping integrals $J_{mn}=J/|m-n|^{\mu}$. It was
argued recently [A. Rodr\'{\i}guez \emph{at al.}, J.\ Phys.\ A: Math.\ Gen.\
{\bf 33}, L161 (2000)] that this model reveals a localization-delocalization
transition with respect to the disorder magnitude provided $1 < \mu < 3/2$. The
transition occurs at one of the band edges (the upper one for $J>0$ and the
lower one for $J<0$). The states at the other band edge are always localized,
which hints on the existence of a single mobility edge. We analyze the mobility
edge and show that, although the number of delocalized states tends to infinity,
they form a set of null measure in the thermodynamic limit, i.e. the mobility
edge tends to the band edge. The critical magnitude of disorder for the band
edge states is computed versus the interaction exponent $\mu$ by making use of
the conjecture on the universality of the normalized participation number
distribution at transition.

\end{abstract}

\pacs{%
71.30.+h;   
72.15.Rn;   
78.30.Ly;   
36.20.Kd    
}

\maketitle

\section{Introduction}
\label{introduction}

In 1958 Anderson formulated a simple tight-binding model with uncorrelated
on-site (diagonal) disorder and predicted a localization-delocalization
transition~(LDT) in three dimensions~(3D): a phase of extended states appears at
the band center in the thermodynamic limit if the disorder magnitude is smaller
than a critical value, while all states are localized at larger magnitudes of
disorder.~\cite{Anderson58} The phase of delocalized states is separated from
the two phases of localized states by two mobility edges.~\cite{Mott67} The
concept of the mobility edge is of key importance for the low-temperature
transport properties of disordered materials.

Since the advent of the single-parameter scaling hypothesis, introduced by
Abrahams~\emph{et al.},~\cite{Abrahams79} the occurrence of a
localization-delocalization transition (LDT) in disordered systems with time
reversal symmetry was ruled out in one-~(1D) and two dimensional~(2D) geometries
at any disorder strength (for the overview see
Refs.~\onlinecite{Lee85,Kramer93,Beenakker97,Janssen98}). The localized nature
of the states in 1D was pointed out even earlier by Mott and Twose.\cite{Mott61}

At the end of eighties and beginning of nineties it was realized however that
correlations in disorder may give rise to extended states in low
dimensions.~\cite{Flores89,Dunlap90,Wu91,Phillips91,Adame1,Adame2,Bellani99,%
Moura98,Izrailev99,Kuhl00,Izrailev01,Moura02,Pouthier02,Izrailev03,Liu03,%
Carpena04} Thus, short-range correlated on-site disorder was found to cause the
appearance of extended states at special resonance energies in 1D. They form a
set of null measure in the density of states in the thermodynamic
limit,~\cite{Flores89,Dunlap90,Wu91,Phillips91,Adame1,Adame2} implying the
absence of the mobility edges in those models. In spite of this fact, even the
infinitesimal fraction of the extended states may have a strong impact on the
transport properties of disordered materials. In particular, short-range
correlations in disordered potential were put forward to explain unusual
conducting properties of polymers, such as polyaniline and heavily doped
polyacetylene,~\cite{Wu91,Phillips91} as well as semiconductor superlattices
grown with random but correlated quantum well sequences.~\cite{Bellani99}

Contrary to short-range correlations in the disorder distribution, long-range
correlations were demonstrated to cause the LDT in~1D systems, which is
analogous to the standard Anderson LDT in
3D.~\cite{Moura98,Izrailev99,Kuhl00,Izrailev01,Moura02,Izrailev03} In this
regard, a 1D system with nearest-neighbor interactions and long-range correlated
on-site disorder distribution with a power-like spectrum $S(k)\sim k^{-\alpha}$
is critical with respect to the exponent $\alpha$. More specifically, when the
standard deviation of the energy distribution equals the nearest-neighbor
hopping and $\alpha < 2$ all states are localized, while for $\alpha > 2$ a
phase of extended states appears at the center of the band giving rise to two
mobility edges. The phase occupies a finite fraction of the density of states. A
similar picture holds in 2D.~\cite{Liu03} The authors of
Ref.~\onlinecite{Kuhl00} proposed to use the long-range correlated disorder and
the appearance of a phase of extended states for designing microwave filters
based on a single-mode waveguide. This type of disorder is also being studied in
biophysics in connection with the large-distance charge transport in DNA
sequences.~\cite{Carpena02,Yamada04}

Another 1D model which exhibits the LDT and a phase of extended states is an
ensemble of power-law random banded matrices $H_{ik} \propto G_{ik}\,|i
-k|^{-\alpha}$, where the matrix $G_{ik}$ runs over a Gaussian orthogonal
ensemble~\cite{Levitov89,Mirlin96} (for an overview see
Ref.~\onlinecite{Mirlin00a}). This model is critical \emph{with respect to the
interaction exponent\/} $\alpha$: for $\alpha > 1$ all states are localized,
while all of them are delocalized at $\alpha < 1$, suggesting that $\alpha = 1$
is the critical point in the model. Within the framework of this model, it was
demonstrated rigorously that (i)~the distribution function of the inverse
participation ratio is scale invariant at transition and (ii)~the relative
fluctuation of the  inverse participation ratio (the ratio of the standard
deviation to the mean) is of the order of unity at the critical
point.~\cite{Evers00,Mirlin00b} This finding confirmed the conjecture (which was
put forward earlier~\cite{Shapiro86,Cohen88}),  that distributions of relevant
physical magnitudes are universal at criticality  (see also
Refs.~\onlinecite{Shklovskii93,Fyodorov95,Prigodin98}). The invariance can
therefore be used to monitor the critical point.~\cite{Malyshev04a,Malyshev04b}

Recently, several reports addressed to the unusual localization properties of~1D
and~2D tight-binding models with uncorrelated diagonal disorder and
\emph{non-random\/} long-range coupling between sites $\bf m$ and $\bf n$, which
falls according to a power law $J/|\bf m-n|^\mu$,~\cite{Malyshev04a,Malyshev04b,%
Cressoni98,Rodriguez00,Rodriguez03,Balagurov04} (see also
Refs.~\onlinecite{Xiong03}-\onlinecite{Brito04}). More specifically, the states
at one of the band edges (the upper one for $J>0$ and the lower one for $J<0$)
undergo the LDT {\it with respect to the disorder strength} $\Delta$ if the
interaction exponent $\mu$ ranges within the interval $1 < \mu < 3d/2$, $d$
being the dimensionality.~\cite{Rodriguez03} In what follows we set $J>0$,  so
that extended states can appear at the upper band-edge. The states at the other
band-edge are strongly localized, no matter how small the disorder magnitude is,
thus, suggesting the existence of a single mobility edge.\cite{Rodriguez00} At
$\mu \ge 3d/2$ all states were found to be localized. The character of
localization, however, turned out to be governed by the interaction exponent
$\mu$. At $\mu = 3d/2$, the upper band-edge states are weakly localized, similar
to those at the center of the band in the standard 2D Anderson
model,~\cite{Abrahams79} while at $\mu > 3d/2$ they are strongly localized.

In the present paper we deal with the analysis, both numerical and analytical,
of the localization properties of the latter class of 1D~Hamiltonians. We
calculate the phase diagram of the transition for the upper band-edge states,
i.e., the critical disorder magnitude $\Delta_c$ versus the interaction exponent
$\mu$, and show that $\Delta_c$ vanishes as $(3/2 - \mu)^{2/3}$ when $\mu \to
3/2$, while it diverges as $(\mu -1)^{-1}$ when $\mu \to 1$. Applying the finite
size scaling analysis we study the problem of the mobility edge and show that
the mobility edge approaches to the upper band edge in the thermodynamic limit.
In other words, the fraction of the delocalized states forms a set of null
measure, although their number tends to infinity on increasing the system size
$N$ as $N^{(3/2-\mu)/(2-\mu)}$ (note that the dependence is sub-linear). This
numerical finding is supported by a simple qualitative arguments based on the
comparison of size scaling of two magnitudes: the bare level spacing at the
band-edge and effective disorder \emph{seen\/} by the quasiparticle.

The outline of the paper is as follows. In the next Section, we describe the
model and briefly overview qualitative arguments which brought us to the
conjecture on the existence of the LDT  within the model. In
Sec.~\ref{phase_diagram}, we present the phase diagram of the LDT for the upper
band-edge states, which is calculated on the basis of the statistics of the
participation number. The mobility edge and fraction of the delocalized states
are addressed in Sec.~\ref{fraction}. We conclude  the paper in
Section~\ref{summary}.

\section{Model and qualitative reasoning}
\label{model}

We consider a tight-binding Hamiltonian on a 1D \emph{regular\/} lattice with
$N$ sites
\begin{equation}
   {\cal H} = \sum_{n} \varepsilon_{n} |n \rangle \langle n|
   + \sum_{nm} J_{mn} |m \rangle \langle n| \ .
\label{H}
\end{equation}
Here $|n\rangle$ is the ket vector of a state with on-site energy
$\varepsilon_{n}$. These energies are taken at random and uncorrelated for
different sites and distributed uniformly around zero within the interval
$[-\Delta/2, \Delta/2$], having therefore zero mean,
$\langle\varepsilon_{n}\rangle = 0$, and standard deviation 
$\langle\varepsilon_{n}^2\rangle^{1/2} = \Delta/\sqrt{12}$ (the angle brackets
$\langle \cdots \rangle$ denote average over disorder realizations). The hopping
integrals $J_{mn}$ do not fluctuate and are set in the form
$J_{mn}=J/|m-n|^{\mu}$, with $J > 0 $ and $J_{nn}=0$.

First, we address the disorder-free system ($\Delta = 0$), taking periodic
boundary conditions for the sake of simplicity. Then the eigenstates of the
Hamiltonian~(\ref{H}) are plane waves with quasi-momenta $K = 2\pi k/N$ within
the first Brillouin zone $k \in [-N/2,N/2)$. The corresponding eigenenergies
are given by
\begin{equation}
\label{EnergyK} E_\mu(K) = 2J \sum_{n > 0} \frac{\cos(Kn)}{|n|^\mu}\ ,
\end{equation}
where the summation runs over all $N$ sites of the lattice, and $\mu$ is assumed
to be larger than unity to get a bounded energy spectrum. The complete account
for all terms in the sum~(\ref{EnergyK}) is important in the neighborhood of the
upper band-edge, where the long-range hopping terms affect the dispersion
vastly. At the upper band-edge ($K \to 0$) the dispersion
relation~(\ref{EnergyK}) is as follows~\cite{Balagurov04}
\begin{equation}
\label{top}
    E^\mathrm{top}_\mu(K) = E_\mathrm{top}(\mu) -
    J\,A_\mathrm{top}(\mu) |K|^{\mu - 1} - J\,B_\mathrm{top}(\mu) K^2 \ ,
\end{equation}
when $\mu\neq 3$. Here, $E_\mathrm{top}(\mu) = 2J \sum^\infty_n n^{-\mu}=2J
\zeta(\mu)$ is the upper band-edge energy in the thermodynamic limit,
$A_\mathrm{top}(\mu)$ and $B_\mathrm{top}(\mu)$ are dimensionless positive
constants on the order of unity (for brevity we do not provide explicit
expressions for them), and $\zeta(\mu)$ is the Riemann $\zeta$-function. For
small $K$, the sub-quadratic term in the right-hand side of Eq.~(\ref{top})
dominates over the  quadratic one if $\mu < 3$ and {\it vice versa}. The range
$\mu < 3$ will be of our primary interest. We will focus later on the size
scaling of the energy spacing at the upper band edge at $\mu < 3$, which is
\begin{equation}
\label{topscaling}
    \delta E^\mathrm{top}_\mu \propto  N^{1 - \mu} \ .
\end{equation}

\noindent At the lower band edge $(|K| \to \pi)$, the energy
spectrum is parabolic:
\begin{equation}
\label{bottom}
    E^\mathrm{bot}_\mu(K)  =  E_\mathrm{bot}(\mu) +
    J\,B_\mathrm{bot}(\mu)\,(\pi - K)^2 \ ,
\end{equation}
where $E_\mathrm{bot}(\mu) = 2\,J \sum^\infty_n (-1)^n n^{-\mu}$ is the lower
band energy, which depends weakly on $\mu$, and $B_\mathrm{bot}(\mu)$ is a
dimensionless constant on the order of unity. Respectively, the energy spacing
at the bottom of the band scales as
\begin{equation}
\label{bottomscaling}
    \delta E^\mathrm{bot}_\mu \propto  N^{-2} \ .
\end{equation}

On introducing the disorder the eigenstates of the regular system couple to each
other, which can result in their localization. The typical fluctuation of the
coupling matrix, namely the first term on the right-hand side of Eq.~(\ref{H})
in the $K$-space basis, is~\cite{Rodriguez00}
\begin{equation}
\label{sigma}
    \sigma  = \frac{\Delta}{\sqrt{12N}} \ .
\end{equation}
Now, we compare the size dependence of $\sigma$ to that of the eigenenergy
spacing at the top of the band $\delta E^\mathrm{top}_\mu$ (defined by
Eq.~(\ref{topscaling})). As the system size increases the typical value of the
eigenstate coupling $\sigma$ decreases faster than the energy  spacing $\delta
E^\mathrm{top}_\mu$ at $\mu < 3/2$.  Consider a system of size $N$ and let
$\sigma$ be of \emph{perturbative\/}  magnitude (i.e. $\sigma \ll \delta E$),
then the upper band-edge states are only weakly perturbed by disorder and remain
extended over the whole system. Upon increasing the system size, the inequality
$\sigma \ll \delta E$ gets even stronger, so that the perturbation becomes
weaker and therefore the upper states will remain delocalized at $N\to\infty$.
On the other hand, large disorder  (say, larger than the bare bandwidth) would
certainly localize all the states. These arguments indicate that there can exist
extended states at the upper band-edge at finite disorder, provided that $1 <
\mu < 3/2$. This conjecture was confirmed both theoretically, by means of the
renormalization group approach combined with a super-symmetric method for
disorder averaging~\cite{Rodriguez03}, and
numerically.~\cite{Malyshev04a,Malyshev04b,Rodriguez00} The value $\mu = 3/2$
represents the marginal case in which the upper band-edge states are localized
weakly.~\cite{Malyshev04a,Malyshev04b,Rodriguez03}  In what follows, we focus
therefore on the interaction exponent $\mu$ ranging within the interval $1 < \mu
\leq 3/2$.

At the bottom of the band $(|K| \to \pi)$ the level spacing  diminishes as
$N^{-2}$, that is faster than the effective magnitude of disorder $\sigma$.
Thus, even if $\sigma \ll \delta E$ for some lattice size and the states are
extended over the whole system, the inequality will be reversed for larger $N$,
which will finally result in strong coupling of the states and their eventual
localization. The same conclusion holds for the entire energy spectrum if $\mu >
3/2$.

The above picture implies the existence of a single mobility edge separating the
phases of localized and delocalized states. We address this question in detail
in Sec.~\ref{fraction}.
\begin{figure}[ht]
\includegraphics[width=0.75\columnwidth,clip]{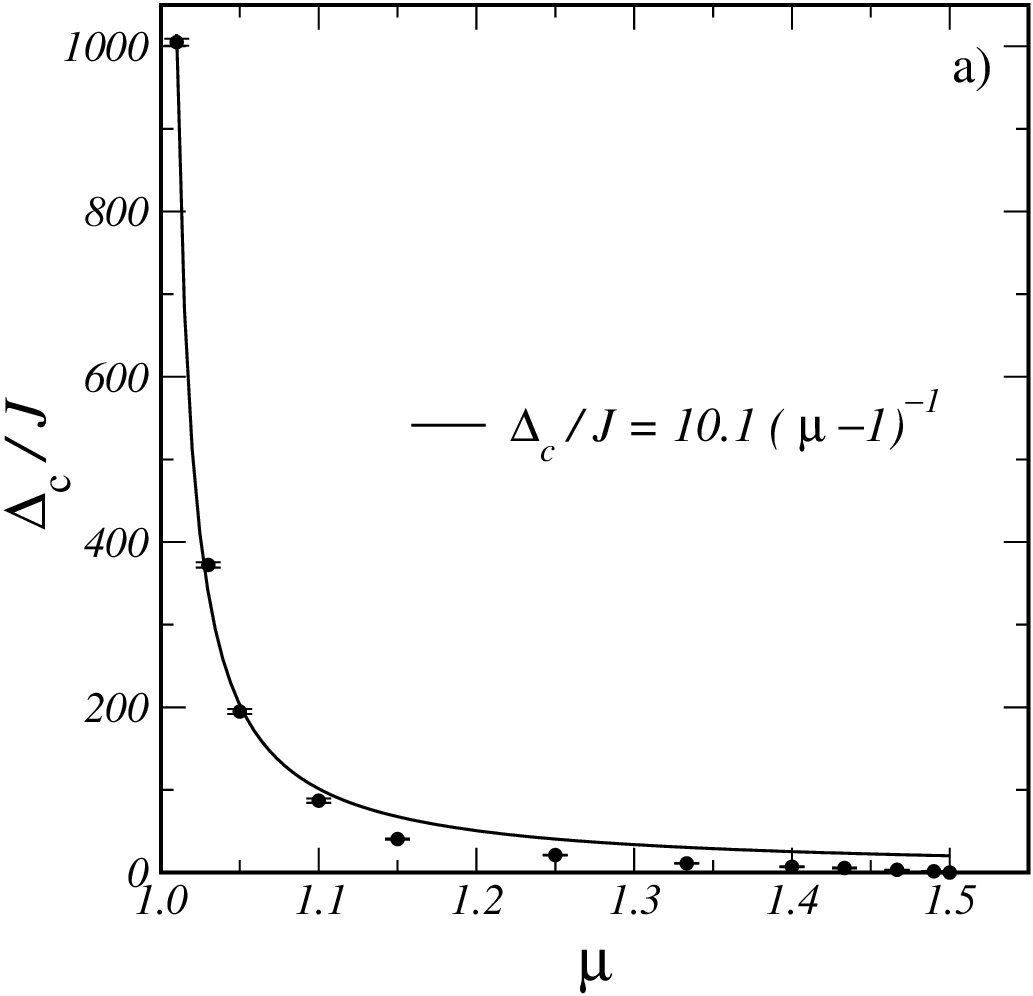}
\includegraphics[width=0.75\columnwidth,clip]{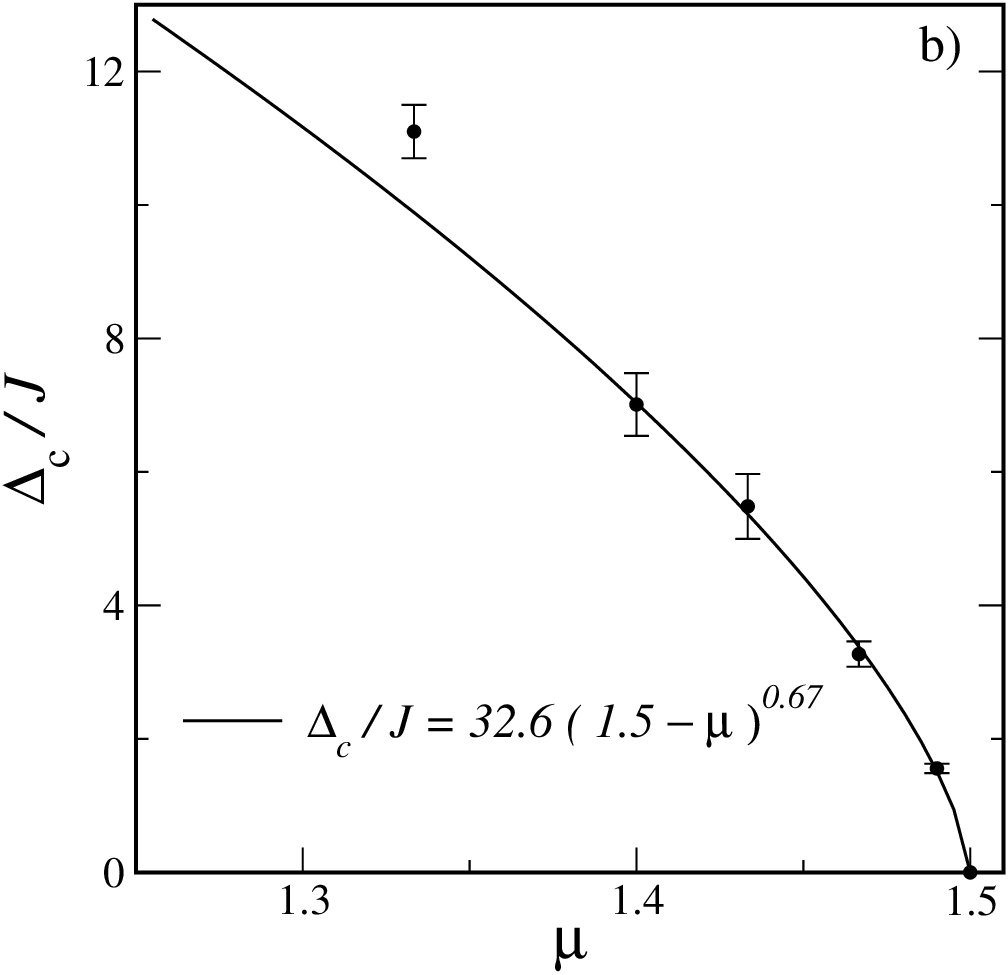}
\caption{
    Critical magnitude of disorder $\Delta_c$ as a function of
    the interaction exponent $\mu$. Full circles represent the
    numerical data, while solid lines are the best fits:
    ({\it a}) \, $\mu \to 1$ and ({\it b}) \,  $\mu \to 3/2$.
    }
    \label{fig1}
\end{figure}

\section{Phase diagram of the transition}
\label{phase_diagram}

In this Section we calculate the dependence of critical disorder magnitude
$\Delta_c$ on the interaction exponent $\mu \in (1,3/2]$. To detect the
transition we analyze the wave function statistics. More specifically, we
calculate size and disorder dependencies of the relative fluctuation of the
participation number~(PN) defined as
\begin{equation}
    {P_k}=\left(\sum_{n=1}^{N} |\psi_{k n}|^{4}\right)^{-1} \ ,
\label{PN}
\end{equation}
where $\psi_{k n}$ denotes the $n$-th component of the normalized $k$-th
eigenstate of the Hamiltonian~(\ref{H}). Here, the state index $k$ ranges
from $1$ to $N$. By definition, we adscribe band-edge states $k=1$ and $k=N$ to
the uppermost and lowermost eigenstates, respectively. It was demonstrated
recently~\cite{Malyshev04a,Malyshev04b} that within the considered model, the
ratio of the standard deviation of the PN (SDPN) to the mean of the PN (MPN) is
an invariant parameter at transition and therefore can be used to detect the
critical point. We note that wave function statistics turned out to be more
efficient for this purpose than level statistics (the latter also represents a
method to monitor the transition~\cite{Shklovskii93}), because wave function
statistics  appeared to be less affected by strong finite size effects  within
the model (for details see the discussion in
Refs.~\onlinecite{Malyshev04a}-\onlinecite{Malyshev04b}).

As the LDT occurs at the top of the band within the considered model, we
calculated disorder and size dependencies of the ratio SDPN/MPN for the
uppermost state. Open chains were used in all calculations.  Following the
procedure developed in Ref.~\onlinecite{Malyshev04b}, we obtained the critical
magnitude of disorder $\Delta_c$ at which uppermost eigenstates undergo the LDT
for $\mu \in (1,3/2]$. The results of the simulations are shown in
Fig.~\ref{fig1} by the full circles, while solid lines represent best fits close
to the limiting points $\mu = 1$ (Fig.~\ref{fig1}{\it a}) and $\mu = 3/2$
(Fig.~\ref{fig1}{\it b}). We found that
\begin{subequations}
\begin{align}
    \Delta_c & \approx 10.1 J (\mu-1)^{-1} \ ,
    & \mu \to 1 \ ,
    \label{muto1} \\
    \Delta_c & \approx 32.6J \left(3/2-\mu\right)^{0.67} \ .
    & \mu \to 3/2 \ ,
    \label{muto1.5}
\end{align}
\end{subequations}
First of all, we notice that, according to Eq.~(\ref{muto1}), the critical
magnitude of disorder diverges as $\Delta_c \propto (\mu-1)^{-1}$ when $\mu \to
1$. The explanation of the divergence relies in the fact that $\Delta_c \propto
E^\mathrm{top}_\mu$ for $\Delta_c\gg 1$,~\cite{Malyshev04b} and the upper
band-edge energy $E^\mathrm{top}_{\mu=1}$ diverges when $N \to \infty$ as
$(\mu-1)^{-1}$. Contrary to that, the critical magnitude of disorder vanishes as
$\Delta_c \propto (3/2-\mu)^{0.67}$ when $\mu \to 3/2$, indicating that in the
marginal case ($\mu = 3/2$) the uppermost state is localized ($\Delta_c=0$), in
full agreement with the results obtained by the renormalization group approach
combined with a super-symmetric method for disorder
averaging.~\cite{Rodriguez03}

\section{Mobility edge and fraction of the delocalized states}
\label{fraction}

Having discussed the dependence of the critical magnitude of disorder $\Delta_c$
on the interaction exponent $\mu$ for the uppermost state, we now focus on the
mobility edge and  the fraction of extended states in the thermodynamic limit.
This question has not been addressed in previous
studies,~\cite{Malyshev04a,Malyshev04b,Cressoni98,Rodriguez00,
Rodriguez03,Balagurov04} except for a comment on the existence of a single
mobility edge~\cite{Rodriguez00, Rodriguez03} (see also the discussion in
Sec.~\ref{model}).

\begin{figure}[ht!]
\includegraphics[width=0.75\columnwidth,clip]{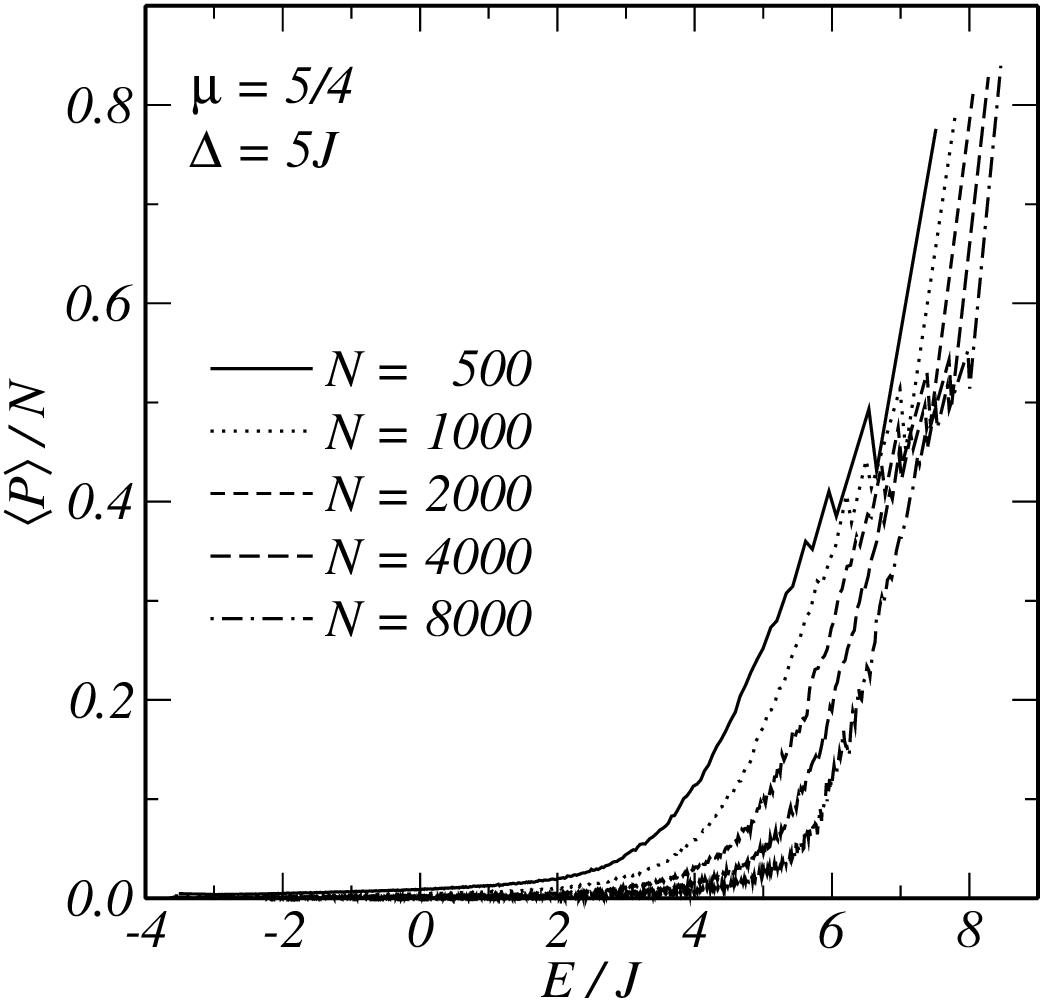}
\caption{
    The normalized mean participation number $\langle P_k\rangle/N$ as a
    function of energy for different system sizes $N$ calculated for the
    interaction exponent $\mu = 5/4$ and the disorder strength $\Delta = 5J$.
    }
    \label{fig2}
\end{figure}

To work out this problem, we numerically diagonalized the Hamiltonian~(\ref{H})
and calculated the normalized MPN, $\langle P\rangle/N$, as a function of energy
for different system sizes $N$. The results of these simulations are shown in
Fig.~\ref{fig2} for a particular set of interaction exponent and  magnitude of
disorder ($\mu=5/4$ and $\Delta=5J$). First, one can see that the upper
band-edge energy increases with $N$ as was mentioned in Sec.~\ref{phase_diagram}
(see also Refs.~\onlinecite{Malyshev04a}-\onlinecite{Malyshev04b}). Note also
the noticeable size dependence of the normalized MPN of the uppermost state.
This behavior reflects the finite size effects already mentioned above (see the
preceding section). The boundaries result in a positive correction of the order
of $N^{-\mu}$ to the \emph{bulk\/} value  of the uppermost wave function
($\sim 1/\sqrt{N}$). Consequently, the normalized MPN depends on the system size as
$(1 - c\,N^{1-\mu})$, where $c$ is a constant. The second, and the most
important, observation is that the normalized MPN increases monotonically on
approaching the upper band edge, for all considered values of $N$.

\begin{figure}[ht]
\includegraphics[width=0.75\columnwidth,clip]{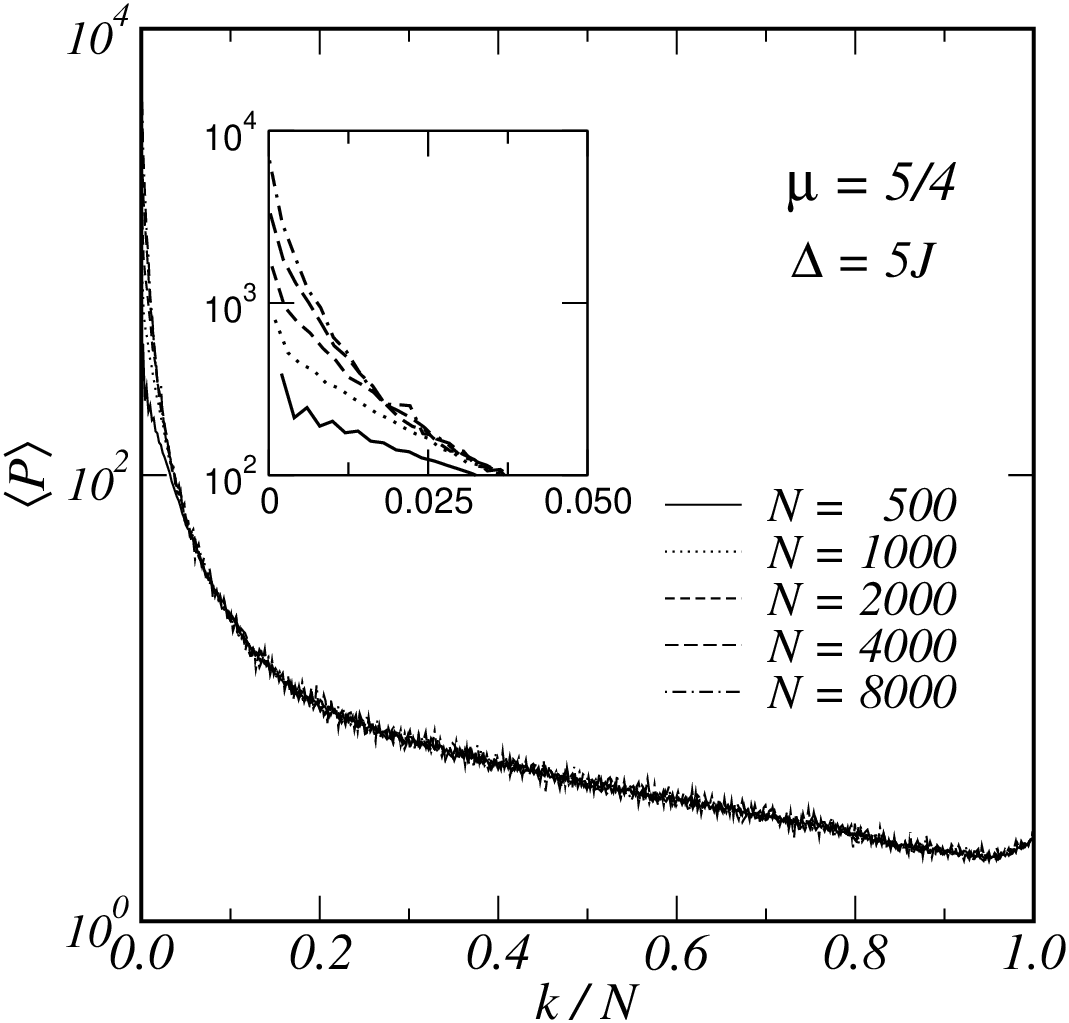}
\caption{
    The mean participation number as a function of the normalized state index
    $k$ for different system sizes $N$ calculated for the interaction exponent
    $\mu = 5/4$ and the disorder strength $\Delta = 5J$. The inset shows an
    enlarged view close to the upper band edge.
    }
    \label{fig3}
\end{figure}

In order to avoid the size dependence of the upper band-edge, we use hereafter
the state index $k$ rather than the energy $E_\mu(k)$. In Fig.~\ref{fig3} we
plotted the MPN as a function of the normalized state index $k/N$. The figure
demonstrates that the MPN is independent of the system size in a wide range of
the normalized state index $k/N$. The perfect collapse of the curves within this
range clearly indicates the localized nature of these eigenstates. However, at
the top of the band the MPN increases linearly with the system size and the
collapse is absent (see the blow-up in the inset of Fig.~\ref{fig3}). This
result suggests that not only the uppermost eigenstate, but a number of them are
delocalized, in agreement with the previous claim raised in
Ref.~\onlinecite{Rodriguez03}

\begin{figure}[ht]
\includegraphics[width=0.75\columnwidth,clip]{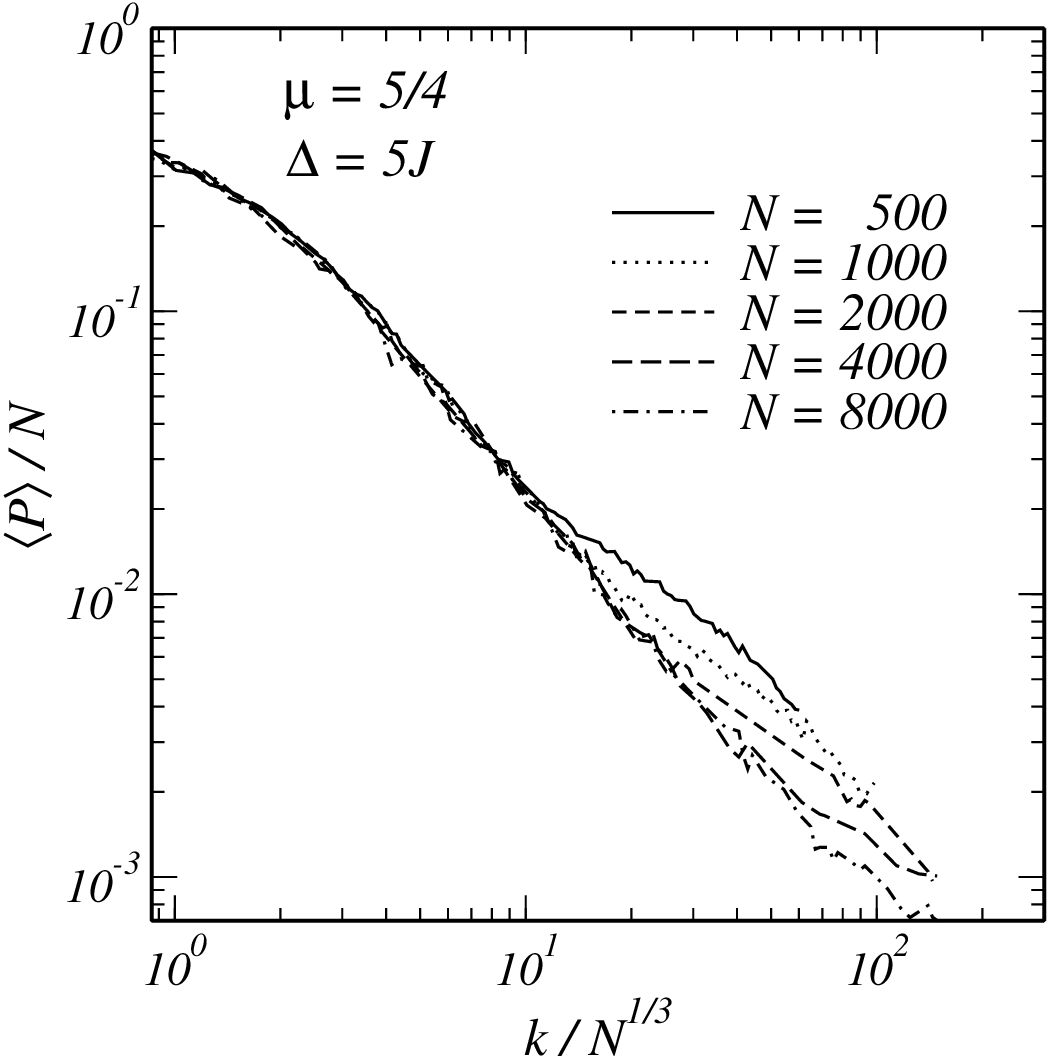}
\caption{
    Normalized mean participation number as a function of $k/N^{1/3}$ for
    different system sizes calculated for $\mu=5/4$ and $\Delta=5J$.
    }
    \label{fig4}
\end{figure}

We now apply a finite size analysis of our numerical results to further confirm
the latter statement and to obtain the size scaling of the number of extended
eigenstates. Figure~\ref{fig4} shows that the MPN for different sizes collapses
onto a single curve close to the top of the band after introducing the rescaled
index $k/N^{1/3}$ for $\mu=5/4$. The collapse holds up to a finite value of the
rescaled index, which indicates that the number of extended states scales as
$\propto N^{1/3}$ when $\mu=5/4$.

To provide support to these numerical results, we now develop a simple
(perturbative) analytical approach that allows us to understand the origin of
the obtained results. To this end, we analyze the energy spacing close to the
upper band-edge in the disorder-free system ($\Delta = 0$) more accurately. The
spacing can be obtained from Eq.~(\ref{top}):
\begin{equation}
\label{dE1}
    {\delta E}^\mathrm{top}_\mu(k) = J\,C_\mu \> \frac{1}{k}
    \left(\frac{k}{N}\right)^{\mu - 1} \ .
\end{equation}
The constant $C_\mu$ absorbs all unessential numerical factors. Note the absence
of the scaling of the spacing $\delta E^\mathrm{top}_\mu(k)$, i.e., the spacing
depends on $k$ not only on $k/N$. This is crucial for the peculiar features of
the LDT within the model. Indeed, consider a chain of a particular size $N$.
Assume that the disorder is perturbative for the uppermost state ($k = 1$),
i.e., $\sigma \ll \delta E^\mathrm{top}_\mu(1)$, so that states at the top  are
extended over the whole sample. Let us now find the \emph{mobility edge\/} for
this finite lattice, defining it by the equality
\begin{equation}
\delta E^\mathrm{top}_\mu(k_m) = \sigma\ .
\label{Ekm}
\end{equation}
The number $k_m$, which is $N$-dependent, divides all eigenstates into two sets:
those with $k < k_m$ are delocalized in the above sense, while eigenstates with
$k > k_m$ are localized in the usual sense. Thus, $k_m$ provides us with the
number of extended  states for a particular system size $N$. Determined by the
equation~(\ref{Ekm}) it reads
\begin{equation}
\label{kc}
    k_m=D_\mu\left(\frac{J}{\Delta}\right)^{1/(2-\mu)}N^{(3/2-\mu)/(2-\mu)}\ ,
\end{equation}
where all unessential constants are absorbed into $D_\mu$. Applying
Eq.~(\ref{kc}) to the particular case $\mu=5/4$ we recover the behavior that we
have found numerically,  namely $k_m \propto N^{1/3}$.

\begin{figure}[ht]
\includegraphics[width=0.75\columnwidth,clip]{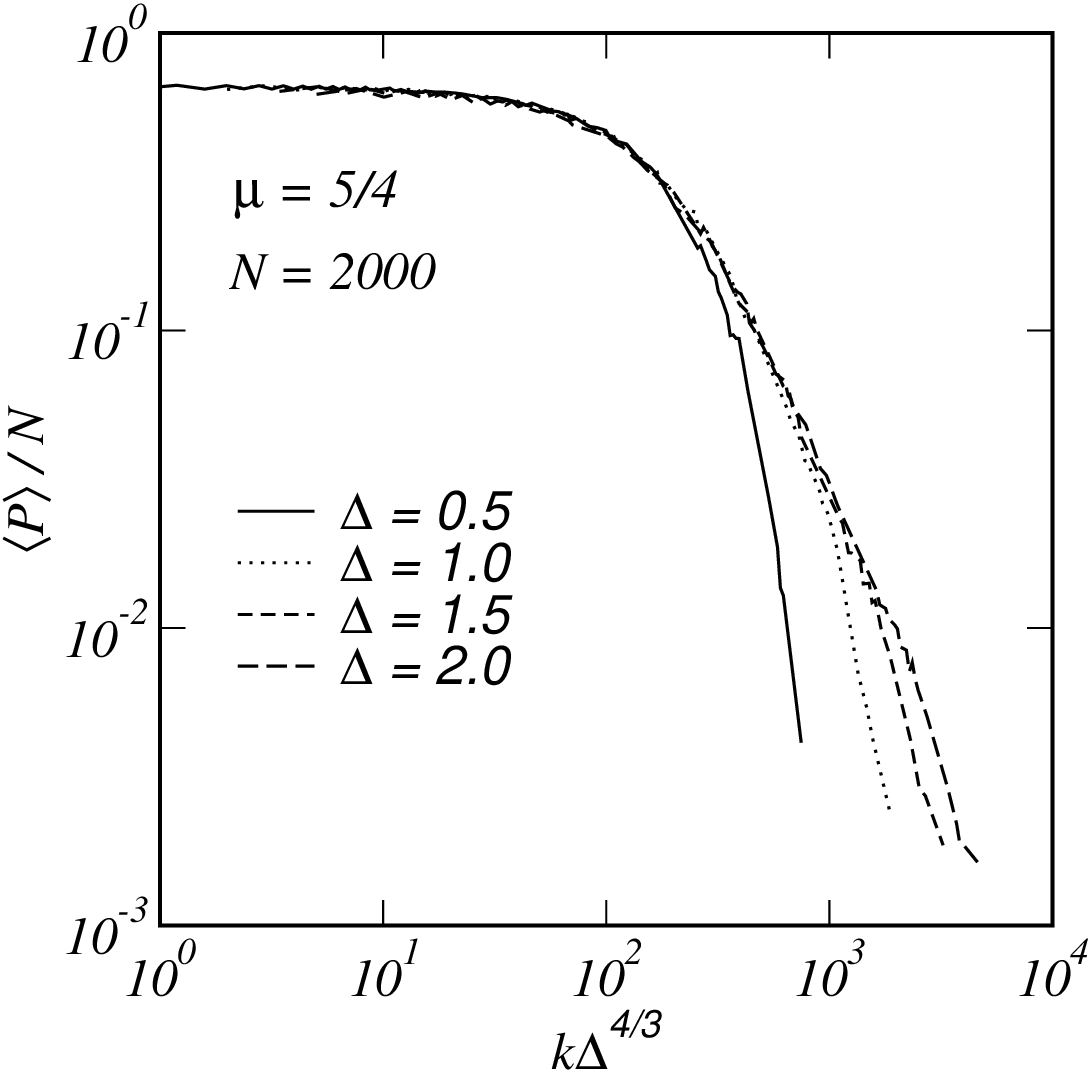}
\caption{
    Normalized mean participation number as a function of
    $k\Delta^{4/3}$ for different disorder magnitudes with
    $\mu=5/4$ and $N=2000$.
    }
    \label{fig5}
\end{figure}

The relationship~(\ref{kc}) provides us also with the dependence of $k_m$ on the
disorder magnitude $\Delta$, which can also be compared to numerical
calculations. We performed such comparison for the interaction exponent $\mu =
5/4$. In this particular case, the exponent $1/(2 - \mu)$ in the
$\Delta$-dependence of $k_m$ is equal to $4/3$. Figure~\ref{fig5} shows the
normalized mean participation number as a function of $k\Delta^{4/3}$ calculated
for different disorder strength $\Delta$ and a given system size $N=4000$. The
collapse of all curves onto a single one in the vicinity of the upper band-edge
supports the validity of Eq.~(\ref{kc}).

We have also tested Eq.~(\ref{kc}) against numerical simulations performed for
other values of $\mu \in (1,3/2)$ and always found an excellent agreement of the
numerical data with the analytic formula~(\ref{kc}). It follows from the formula
that the number of extended states increases sub-linearly with the system size,
namely as $N^{(3/2-\mu)/(2-\mu)}$, which means that the fraction of these states
vanishes as $N^{-1/(4-2\mu)}$ when $N \to \infty$. From this we conclude that
the mobility edge approaches to the upper band edge in the thermodynamic limit.

\section{Summary and concluding remarks}
\label{summary}

We studied both analytically and numerically the localization properties of the
1D tight-binding model with diagonal disorder and \emph{non-random\/} long-range
inter-site interactions, $J_{mn} = J/|{m-n}|^{\mu}$ where $J > 0$. The model can
be critical at the upper band-edge provided $1 < \mu < 3/2$.

We calculated the phase diagram of the transition (the dependence of the
critical magnitude of disorder $\Delta_c$ on the interaction exponent $\mu$) by
studying the participation number statistics. The critical magnitude of disorder
was detected using the size invariance of the ratio of the standard deviation to
the mean. We found that $\Delta_c$ diverges as $(\mu - 1)^{-1}$ when $\mu \to
1$, which originates from the similar divergence of the upper band-edge energy.
If $\mu \to 3/2$, the critical disorder magnitude vanishes as $(3/2 -
\mu)^{2/3}$, indicating that all states are localized at $\mu = 3/2$, no matter
how small the disorder is.

It is shown, both analytically and by means of the finite size scaling analysis,
that the number of extended states at the upper band-edge increases sub-linearly
with the system size $N$, namely  as $\propto N^{(3/2-\mu)/(2-\mu)}$), therefore
forming a set of null measure in the density of states in the thermodynamic
limit. This suggests that the mobility edge is only a meaningful concept for a
finite size system; it approaches to the upper band-edge in the thermodynamic
limit. Although extended states form an infinitesimal fraction of the whole
density of states, they can provide a strong impact on the transport properties,
similarly to what happens in systems with correlated
randomness.~\cite{Wu91,Phillips91,Bellani99}

To conclude, we note that our findings, apart from being interesting from the
theoretical point of view, are relevant for real physical systems. Thus, some
organic materials with planar geometry, in which optically allowed excitations
are dipolar Frenkel
excitons,~\cite{Nabetani95,Tomioka96,Fukumoto98,Daehne99,Dominguez-Adame00,%
Vitukhnovsky00,Vitukhnovsky01,Bakalis02} represent an example; the value of the
interaction exponent for the dipole-dipole inter-site interaction is $\mu = 3$,
which resembles the weak localization regime in the 1D long-range model when
$\mu = 3/2$.~\cite{Rodriguez03} Dipole-exchange spin waves in ferromagnetic
films provide yet another example where these results are
relevant.~\cite{Costa00}

\begin{acknowledgments}

Work at Macei\'{o} was partially supported by the Brazilian research agencies
CNPq and CAPES as well as by the Alagoas state research agency FAPEAL. Work at
Madrid was supported by MCyT (MAT2003-01533) and CAM (Project GR/MAT/0039/2004).
V.\ A.\ M. acknowledges support from ISTC (grant \#2679) and A.\ V.\ M. from
INTAS (YSF 03-55-1545). 

\end{acknowledgments}


\begin{thebibliography}{99}

\bibitem{Anderson58} P.\ W.\ Anderson, Phys.\ Rev.\ \textbf{109},
    1492 (1958).

\bibitem{Mott67}N.\ F.\ Mott, Adv. Phys. \textbf{16}, 49 (1967).

\bibitem{Abrahams79} E.\ Abrahams, P.\ W.\ Anderson, D.\ C.\
    Licciardello, and T.\ V.\ Ramakrishnan, Phys.\ Rev.\ Lett.\
    \textbf{42}, 673 (1979).

\bibitem{Lee85} P.\ A.\ Lee and T.\ V.\ Ramakrishnan, Rev.\ Mod.\
    Phys.\ {\bf 57}, 287 (1985).

\bibitem{Kramer93} B.\ Kramer and A.\ MacKinnon, Rep.\ Prog.\ Phys.\
    {\bf 56}, 1469 (1993).

\bibitem{Beenakker97} C.\ W.\ J.\ Beenakker, Rev. Mod. Phys. {\bf
    69}, 731 (1997).

\bibitem{Janssen98} M. Janssen, Phys. Rep. {\bf 295}, 1 (1998).

\bibitem{Mott61}N.\ F.\ Mott and W.\ D.\ Twose, Adv. Phys. {\bf 10},
    107 (1961).

\bibitem{Flores89} J.\ C.\ Flores, J.\ Phys.:\ Condens.\ Matter
    {\bf 1}, 8471 (1989); J.\ C.\ Flores and M. Hilke J.\ Phys. A:\
    Math.\ Gen. {\bf 26}, L1255 (1993).

\bibitem{Dunlap90} D.\ H.\ Dunlap, H.-L.\ Wu, and P.\ W.\ Phillips,
    Phys.\ Rev.\ Lett.\ {\bf 65}, 88 (1990).

\bibitem{Wu91} H.-L.\ Wu, and P.\ Phillips, Phys.\ Rev.\ Lett.\
    {\bf 66}, 1366 (1991).

\bibitem{Phillips91} P.\ W.\ Phillips and H.-L.\ Wu, Science
    \textbf{252}, 1805 (1991).

\bibitem{Adame1} A.\ S\'{a}nchez and F.\ Dom\'{\i}nguez-Adame, J.
    Phys A: Math. Gen. \textbf{27}, 3725 (1994); A.\ S\'{a}nchez, E.\
    Maci\'{a}, and F.\ Dom\'{\i}nguez-Adame, Phys. Rev. B \textbf{49},
    147 (1994).

\bibitem{Adame2} E.\ Diez, A.\ S\'{a}nchez, F.\ Dom\'{\i}nguez-Adame,
    Phys\ Rev.\ B \textbf{50}, 14359 (1994); F.\ Dom\'{\i}nguez-Adame,
    E. Diez, and A.\ S\'{a}nchez, Phys.\ Rev.\ B \textbf{51}, 8115
    (1995).

\bibitem{Bellani99} V.\ Bellani, E.\ Diez, R.\ Hey, L.\ Toni, L.\
    Tarricone, G.\ B.\ Parravicini, F.\ Dom\'{\i}nguez-Adame, and R.\
    G\'{o}mez-Alcal\'{a}, Phys.\ Rev.\ Lett.\ {\bf 82}, 2159 (1999).

\bibitem{Moura98} F.\ A.\ B.\ F.\ de Moura and M.\ L.\ Lyra, Phys.\
    Rev.\ Lett.\ {\bf 81}, 3735 (1998); Physica A {\bf 266}, 465
    (1999).

\bibitem{Izrailev99} F.\ M.\ Izrailev and A.\ A.\ Krokhin, Phys.\
    Rev.\ Lett.\ {\bf 82}, 4062 (1999).

\bibitem{Kuhl00} U.\ Kuhl, F.\ M.\ Izrailev, A.\ A.\ Krokhin, and
    H.-J.\ St\"{o}ck\-mann, Appl.\ Phys.\ Lett.\ {\bf 77}, 633 (2000).

\bibitem{Izrailev01} F.\ M.\ Izrailev, A.\ A.\ Krokhin, and S.\ E.\
    Ulloa, Phys.\ Rev.\ B {\bf 63}, 041102(R) (2001).

\bibitem{Izrailev03} F.\ M.\ Izrailev and  N.\ M.\ Makarov, Phys.\
    Rev.\ B {\bf 67}, 113402 (2003); Appl.\ Phys.\ Lett.\ {\bf 84},
    5150 (2004).

\bibitem{Moura02} F.\ A.\ B.\ F.\ de Moura, M.\ D.\ Coutinho-Filho,
    E.\ P.\ Raposo, and M.\ L.\ Lyra, Phys.\ Rev.\ B {\bf 66},
    014418 (2002); {\it ibid.} {\bf 68}, 012202 (2003).

\bibitem{Pouthier02} V.\ Pouthier and C.\ Girardet, Phys.\ Rev.\ B
    {\bf 66}, 115322 (2002).

\bibitem{Liu03} W.-S.\ Liu, S.\ Y.\ Liu, and X.\ l.\ Lei, Eur.
    Phys. J. B {\bf 33}, 293 (2003).

\bibitem{Carpena04} P.\ Carpena, P.\ Bernaola-Galv\'{a}n, and P.\ Ch.\ Ivanov,
    Phys.\ Rev.\ Lett.\ \textbf{93}, 176804 (2004).

\bibitem{Carpena02}P.\ Carpena, P.\ B.\ Galvan, P.\ Ch.\ Ivanov,
    and H.\ E.\ Stanley, Nature {\bf 418}, 955 (2002); {\it ibid}
    {\bf 421}, 764 (2003).

\bibitem{Yamada04} H.\ Yamada, Int. J. Mod. Phys. B {\bf 18}, 1697
    (2004); Phys. Lett. A {\bf 332}, 65 (2004).

\bibitem{Levitov89} L.\ S.\ Levitov, Europhys.\ Lett.\ \textbf{9},
    83 (1989); Ann.\ Phys.\ (Leipzig) \textbf{8}, 507 (1999).

\bibitem{Mirlin96} A.\ D.\ Mirlin, Y.\ V.\ Fyodorov, F.-M.\ Dittes,
    J.\ Quezada, and T.\ H.\ Seligman, Phys.\ Rev.\ E \textbf{54},
    3221 (1996).

\bibitem{Mirlin00a} A.\ D.\ Mirlin, Phys.\ Rep. \textbf{326}, 259
    (2000).

\bibitem{Evers00} F.\ Evers and A.\ D.\ Mirlin, Phys.\ Rev.\ Lett.
    \textbf{84}, 3690 (2000).

\bibitem{Mirlin00b}  A.\ D.\ Mirlin and F.\ Evers, Phys.\ Rev.\ B
    \textbf{62}, 7920 (2000).

\bibitem{Shapiro86} B.\ Shapiro, Phys.\ Rev.\ B {\bf 34}, R4394
    (1986); Phil.\ Mag.\ B {\bf 56}, 1031 (1987).

\bibitem{Cohen88}A.\ Cohen, Y.\ Roth, and B.\ Shapiro, Phys. Rev. B
    {\bf 38}, 12125 (1988).

\bibitem{Shklovskii93} B.\ I.\ Shklovskii, B.\ Shapiro, B.\ R.\
    Sears, P.\ Lambrianides, and H.\ B.\ Shore, Phys.\ Rev.\ B
    \textbf{47}, 11\,487 (1993).

\bibitem{Fyodorov95} Y.\ V.\ Fyodorov and A.\ D.\ Mirlin, Phys.\
    Rev.\ B \textbf{51}, 13403 (1995).

\bibitem{Prigodin98} V.\ N.\ Prigodin and B.\ L.\ Altshuler,
    Phys.\ Rev.\ Lett. \textbf{80}, 1944 (1998).

\bibitem{Malyshev04a} A.\ V.\ Malyshev, F.\ Dom\'{\i}nguez-Adame,
    and V.\ A.\ Malyshev, Phys. Stat. Sol. (b) {\bf 214}, 2419
    (2004).

\bibitem{Malyshev04b} A.\ V.\ Malyshev, V.\ A.\ Malyshev, and F.\
    Dom\'{\i}nguez-Adame, Phys. Rev. B {\bf 70}, 172202 (2004).

\bibitem{Cressoni98} J.\ C.\ Cressoni and M.\ L.\ Lyra, Physica A
    \textbf{256}, 18 (1998).

\bibitem{Rodriguez00} A.\ Rodr\'{\i}guez, V.\ A.\ Malyshev, and F.\
    Dom\'{\i}nguez-Adame, J.\ Phys.\ A: Math.\ Gen.\ {\bf 33},
    L161 (2000).

\bibitem{Rodriguez03} A.\ Rodr\'{\i}guez, V.\ A.\ Malyshev, G.\
    Sierra, M.\ A.\ Mart\'{\i}n-Delgado, J.\ Rodr\'{\i}guez-Laguna,
    and F.\ Dom\'{\i}nguez-Adame, Phys.\ Rev.\ Lett.\ \textbf{90},
    27404 (2003).

\bibitem{Balagurov04} D.\ B.\ Balagurov, V.\ A.\ Malyshev, and
    F.\ Dom\'{\i}nguez-Adame, Phys.\ Rev.\ B {\bf 69}, 104204
    (2004).

\bibitem{Xiong03} S.-J.\ Xiong and G.-P.\ Zhang, Phys.\ Rev.\ B {\bf 68}, 174201
    (2003).

\bibitem{Brito04} P.\ E.\ de Brito, E.\ S.\ Rodrigues, H.\ N.\ Nazareno,
    Phys.\ Rev.\ B {\bf 69}, 214204 (2004).

\bibitem{Nabetani95} A.\ Nabetani, A.\ Tamioka, H.\ Tamaru, and
    K.\ Miyano, J. Chem. Phys. {\bf 102}, 5109 (1995).

\bibitem{Tomioka96} A.\ Tomioka and K.\ Miyano, Phys.\ Rev.\ B {\bf
    54}, 2963 (1996).

\bibitem{Fukumoto98} H.\ Fukumoto and Y.\ Yonezawa, Thin Solid
    Films {\bf 327-329}, 748 (1998).

\bibitem{Daehne99} L.\ Daehene and E.\ Biller, Phys.\ Chem.\ Chem.\
    Phys. {\bf 1}, 1727 (1999).

\bibitem{Dominguez-Adame00} F.\ Dom\'{\i}nguez-Adame, V.\ A.\
    Malyshev, and A.\ Ro\-dr\'{\i}\-guez, J.\ Chem.\ Phys.\ {\bf 112},
    3023 (2000).

\bibitem{Vitukhnovsky00} A.\ G.\ Vitukhnovsky, A.\ N.\ Lobanov, A.\
    V.\ Pimenov, Y.\ Yonezawa, N.\ Kometani, K.\ Asami, and J.\ Yano,
    J. Lumin. {\bf 87-89}, 260 (2000).

\bibitem{Vitukhnovsky01} A.\ G.\ Vitukhnovsky, A.\ N.\ Lobanov, A.\
    V.\ Pimenov, Y.\ Yonezawa, and N.\ Kometani, Int.\ J.\ Mod.\
    Phys. B {\bf 15}, 4017 (2001).

\bibitem{Bakalis02} L.\ D.\ Bakalis, I.\ Rubtsov, and J.\ Knoester,
    J.\ Chem.\ Phys.\ {\bf 117}, 5393 (2002).

\bibitem{Costa00} R.\ N.\ Costa Filho, M.\ G.\ Cottam, and G.\ A.\ Farias,
    Phys.\ Rev.\ B {\bf 62}, 6545 (2000).

\end{thebibliography}
\end{document}